\documentclass[12pt]{article}
\usepackage[cp1251]{inputenc}
\usepackage[english]{babel}
\usepackage{graphicx,graphics}

 0
\begin{document}

\title{A dressing of zero-range potentials and
electron-molecule scattering problem at Ramsauer-Townsend minimum}
\author{  S.B. Leble
\small \\ Theoretical Physics and Mathematical Methods Department,
\small\\ Gdansk University of Technology, ul, Narutowicza 11/12,
Gdansk, Poland,
\small \\ leble@mifgate.pg.gda.pl \\  \\[2ex]
S. Yalunin \small\\ Theoretical Physics Department, \small\\
Kaliningrad State University, A. Nevsky st. 14, Kaliningrad,
Russia, \small \\ yalunin@bk.ru}

\maketitle
\begin{abstract}
A dressing technique is used to improve zero range potential (ZRP)
model. We consider a Darboux transformation starting with a ZRP,
the result of the "dressing" gives a potential with non-zero range
that depends on a seed solution parameters. Concepts of the
partial waves and partial phases for non-spherical potential are
used in order to perform Darboux transformation. The problem of
scattering on the regular X$_{\hbox{n}}$ and YX$_{\hbox{n}}$
structures is studied. The results of the low-energy
electron-molecule scattering on the dressed ZRPs are illustrated
by model calculation for the configuration and parameters of the
silane ($\hbox{SiH}_4$) molecule. \center{Key words: low-energy
scattering, multiple scattering, Ramsauer-Townsend minimum,
silane, zero range potential.}
\end{abstract}

%\maketitle

\section{Introduction}
The ideas of zero range potential (ZRP) approach were recently
developed $\cite{Balt}, \cite{LY1}, \cite{Der}$ to widen limits of
the traditional treatment by Demkov and Ostrovsky $\cite{DO1975}$
and Albeverio et al. $\cite{AlGa1988}$. The advantage of the
theory is the possibility of obtaining an exact solution of
scattering problem. The ZRP is conventionally represented as the
boundary condition on the wave function at some point.
Alternatively, the ZRP can be represented as pseudopotential
(Breit $\cite{Breit}, \cite{Der}$).

On the other hand, the factorization of a Schr\"{o}dinger operator
allows to construct in natural way exactly solvable potentials
\cite{IH}. The method has the direct link to Darboux
transformations (DT) theory $\cite{MatSall}$. General starting
point of the theory goes up to the Matveev theorem (see e.g.
\cite{Mat}). The transformation can be also defined on the base of
covariance property of the Schr\"{o}dinger equation with respect
to a transformation of wave function and potential energy.

Let us briefly describe the classical method. The DT for
one-dimensional but matrix problem
\begin{equation}\label{E0}
     a_2 \frac{ \partial^2 \psi}{\partial^2 x}+a_1 \frac{ \partial \psi}{\partial
     x} + a_0 \psi = E \psi
\end{equation}
has the form
\begin{equation}\label{DT}
  \psi^{(1)} = \psi_x - \phi_x \phi^{-1} \psi,
\end{equation}
where $\phi_x$ is partial derivative, $s = \phi_x\phi^{-1}$ is
constructed by matrix (operator-valued) solution $\phi$ of the
equation (\ref{E0}) for a generally different eigenvalue E'. For
the matrix case it is enough to take a set of linearly independent
column solutions.

The principal statement (covariance) formally yields
($\partial=\partial/\partial x$)
\begin{equation}\label{E1}
  L^{(1)} \psi^{(1)} = E \psi^{(1)},\hspace{8mm} L^{(1)} = \sum_{n=0}^2 a_n^{(1)}
 \partial^n,
\end{equation}
explicit expressions for $a_n^{(1)}$ are given in \cite{LeZa} for
arbitrary order operator. We would cite here the
 matrix Darboux dressing formulas for
the second order operator.  The coefficient $a_2$  does not
transform, so it is chosen as $a_2 = -1/2$, while the transform
for the second one generally contains the commutator
\begin{equation}\label{a1}
  a_1^{(1)} = a_1 + [a_2,s],
\end{equation}
which, for the given $a_2$ is zero.  Finally, slightly changing
the notations, the transforms are:
\begin{equation}\label{dr2}
 \begin{array}{c}
   \psi^{(1)} = \psi_x - s \psi \\
 a_0^{(1)} = a_0 + a_1' + [a_1,s] + 2a_2s' + a_2's + [a_2,s]s = \\
  a_0 + a_1' - s',
 \end{array}
\end{equation}
where commutator $[a_1,s]=0$, and primed function is the shortcut
for the derivatives by $x$. The functions $ \psi^{(1)}, u^{(1)} $
are often named "dressed" ones, the auxiliary solutions of the
problem (\ref{E0}) are referred as the "prop" functions. An
eigenfunction and a potential from which one starts ($\psi$ and u
here) are called as the "seed" ones. Let us mention that the proof
of the Darboux covariance relation includes a link that in theory
of solitons \cite{MatSall} is named a (general) Miura
transformation that is identically solved by the substitution $"s
= \phi_x\phi^{-1}".$ We, however, do not use this fact, going by
alternative way. Notice, that in the case of the radial
Schr\"{o}dinger equation, $a_1 = -1/r$, that also simplify the
transform (\ref{dr2}).

Darboux formulas in multi-dimensional space could be applied in
the sense of Andrianov, Borisov and Ioffe ideas \cite{ABI}
combined with ones for the radial Schr\"{o}dinger equation
\cite{MatSall,SL}. In the circumstances, DT technique can be used
so as to correct ZRP model.

We attempt to dress the ZRP by means of a special choice of DT in
order to widen possibilities of the ZRP model. DT modifies the
generalized ZRP (GZRP) boundary condition (Section $\ref{S1}$) and
creates a potential with arbitrarily disposed discrete spectrum
levels for any angular momentum $l$. In the section $\ref{S2}$ we
consider partial waves decomposition for a wave function for a
non-spherical potential so as to dress a multi-centered potential,
which includes n ZRPs \cite{LY2}. The problem we consider is
generic: it includes a three-dimensional potential.  In order to
construct DT we consider some aspects of division of differential
operators and use some ideas from \cite{LeZa}. As an important
example, we consider electron scattering by the X$_{\hbox{n}}$ and
YX$_{\hbox{n}}$ structures within the framework of the ZRP model
(Section $\ref{S4}$). In section $\ref{S5}$ we present our
calculations of the corrected by the dressing integral
cross-section for the electron-silane scattering.

\section{Zero range potentials and dressing} \label{S1}

Our observation shows that generalized ZRPs (see $\cite{Balt}$)
are appear as a result of Darboux transformations. In order to
demonstrate it we consider a radial Schr\"{o}dinger equation for
partial wave $\psi_l$ with orbital momentum $l$. The atomic units
are used throughout the present paper $\hbar=m_e=1$
\begin{equation} \label{E}
\left( -\frac{1}{2}\frac{d^2}{dr^2} -\frac{1}{r}
\frac{d}{dr}+\frac{l(l+1)}{2r^2} + u_l-E \right) \psi_l(r) = 0,
\end{equation}
where $u_l$ are potentials for the partial waves with asymptotic
at infinity
\begin{equation} \label{EINF}
\psi_l(r) \sim \frac{\sin(kr-\frac{l\pi}{2}+\delta_l)}{kr}.
\end{equation}
The equation $(\ref{E})$ describes scattering of a particle with
energy E and momentum $k=\sqrt{2E}$. In the absence of the
potential, partial shifts $\delta_l=0$ and partial waves can be
expressed via spherical functions $\psi_l=j_l(kr)$. Well known,
the equations (\ref{E}) are covariant with respect to DT
($\ref{dr2}$) where
\begin{equation}
\begin{array}{l}
a_2=-\frac{1}{2}, \\
a_1=-\frac{1}{r}^{\mathstrut}, \\
a_0=\frac{l(l+1)}{2r^2}^{\mathstrut}+u_l.
\end{array}
\end{equation}
In DT prop function $\phi$ plays an important role because one is
used for calculation of $s$. The function $\phi$ is a solution of
the equation $(\ref{E})$ at a particular value of energy
$E=-\kappa^2/2$, where we assume $\kappa$ is a real number. If
$\kappa$ is a complex number then dressed potential can be a
complex function. In the case $u_l=0$ the solution $\phi$ are
arbitrary combination of linearly independent solutions. Let us
demonstrate that generalized ZRP can be introduced by DT. For our
purpose convenient to use a chain of DTs (Crum formulas
$\cite{Crum}$ with the wave and prop functions multiplied by $r$),
which for our equation look like
\begin{equation} \label{crum}
\psi_l \rightarrow \psi_l^{(1)}=\hbox{const}\cdot
\frac{W(r\psi_l,r\phi_1,\ldots,r\phi_{2l+1})}
{rW(r\phi_1,\ldots,r\phi_{2l+1})},
\end{equation}
\begin{equation} \label{crum1}
u_l \rightarrow u_l^{(1)}=u_l - (\ln
W(r\phi_1,\ldots,r\phi_{2l+1}))'',
\end{equation}
where $W$ is Wronskian, and $\phi_{m}$ are prop functions. The
transformation $(\ref{crum})$ combines the solution $\psi_l$ and
$2l+1$ solutions $\phi_{m}$. The Crum formulas result from the
replacement of a chain of $2l+1$ first order transformations by a
single ($2l+1$)th order transformation, which happens to be more
efficient in practical calculations. In order to obtain ZRP we
start from zero potential and use prop functions
\begin{equation} \label{sms}
\phi_{m}=h_{l}^{(1)}(\kappa_{m}r),
\end{equation}
where $\kappa_{m }$ are solutions of the algebraical equation
$\kappa_{m}^{2l+1}=\imath \alpha_l$. Here we assume $\alpha_l$ is
real number. The explicit form of the spherical functions
$h_l^{(1)}$ and some important properties of $j_l$ are given in
the Appendix. Direct substitution of $(\ref{sms})$ to Wronskian
shows that
\begin{equation}
W(r\phi_1,\ldots,r\phi_{2l+1})={\rm const}.
\end{equation}
It means that dressed potential $u_l^{(1)}(r>0)=0$. The
transformation $(\ref{crum1})$ allows to calculate potential in
range $r>0$. We state that DTs also yield a generalized ZRP at
$r=0$. In order to prove this we perform transformation
$(\ref{crum})$ and show that $\psi_l^{(1)}$ is a solution for a
generalized ZRP. Since potential is equal zero in the region
$r>0$, enough to determine asymptotic of the wave function.
Substituting $\psi_l=j_l(kr)$ to the Crum formulas, we obtain
\begin{equation}
\psi_l^{(1)}=\hbox{const}\cdot
\frac{W(rj_l(kr),r\phi_1,\ldots,r\phi_{2l+1})}
{rW(r\phi_{1}\ldots,r\phi_{2l+1})}\sim
\end{equation}
\begin{equation}
\frac{1}{2{\rm i}}\left\{(-\imath)^l\frac{e^{\imath kr}}{kr}
\frac{\Delta(\imath k,\kappa_1,\ldots,\kappa_{2l+1})}
{\Delta(\kappa_1,\ldots,\kappa_{2l+1})} - \imath^l\frac{
e^{-\imath kr}}{kr} \frac{\Delta(-\imath
k,\kappa_1,\ldots,\kappa_{2l+1})}
{\Delta(\kappa_1,\ldots,\kappa_{2l+1})}\right\},
\end{equation}
where $\Delta$ is Wandermond determinant. Considering one as
product
\begin{equation}
\Delta(\imath k,{\kappa}_1,\ldots, {\kappa}_{2l+1})={\rm
const}\cdot \prod_{m=1}^{2l+1}(\kappa_m-\imath k)
\end{equation}
we obtain an asymptotic, which coincides with asymptotic of the
solution:
\begin{equation} \label{Smat}
\psi_l^{(1)}=\hbox{const}\cdot(h_l^{(1)}(kr)e^{2\imath \delta_l}-
h_l^{(2)}(kr)),\hspace{7mm} \exp(2\imath
\delta_l)=\prod_{m=1}^{2l+1} \frac{\kappa_m-\imath
k}{\kappa_m+\imath k}.
\end{equation}
It easy to show that wave function $(\ref{Smat})$ describes a
scattering by generalized ZRP with orbital momentum $l$. One is
conventionally represented as the boundary condition at $r=0$ on
the wavefunction. It can be verified by direct substitution to the
boundary condition for generalized ZRP:
\begin{equation} \label{Bound}
\left. \frac{(\frac{d}{dr})^{2l+1}\,
r^{l+1}\psi_l^{(1)}}{r^{l+1}\psi_l^{(1)}}\right|_{r=0}= -\frac{2^l l! \alpha_l
}{(2l-1)!!},
\end{equation}
where $\alpha_l$ -- inverse scattering for partial wave with
orbital momentum $l$. Recall that at low-energies $\tan(\delta_l)
\sim -a_l k^{2l+1}$ for a short-range potential, where $a_l$ is
scattering length. In special case of $l=0$ we obtain $(\ln
r\psi)'=-\alpha$. This generalized boundary condition can be
obtained from asymptotic of the wave function in the vicinity of
zero, which was used some authors $\cite{Balt}$. Let us consider
the scattering matrix on the complex $k$-plane. Each element
$\exp(2\imath \delta_l)$ has $2l+1$ poles at the points $k=\imath
\kappa_m$, which lie on circle on the complex plane. Since the
bound states correspond to the poles on the imaginary positive
semi-axis on complex $k$-plane, bound state exists only if
$\alpha_l>0$ and $l$ is odd number or if $\alpha_l<0$ end $l$ is
even. Otherwise ZRP has an antibound state.

There is some "generalization" of ZRP theory,
then inverse scattering length in original boundary condition is replaced by
"$-k\hbox{cot}( \delta)$" (in case $l$=0). In this model, ZRP may has
two (or more) bound states with non-orthogonal wave functions.
This trouble does not appears in our model because our potential has only one bound state.
However, we note that generalized ZRP has other trouble: bound state with orbital
momentum $l>0$ does not belongs to $L^2$ (the zero range effect).
But we think that one problem does not
fatal because this model reasonably describes a low energy scattering.

{\bf Example 1}: There is a simple example which proves the our
observation: ZRP can be introduced by Darboux transformation. Let
us consider transformation for regular solution $\psi=\sin(kr)/r$
with the prop function $\phi=\exp(\alpha r)/r$. Direct calculation
yields  the wave function
\begin{equation} \label{sss}
\psi^{(1)}=\hbox{const}\cdot(\psi'-s\psi)=\frac{\sin(kr-\arctan(k/\alpha))}{kr},
\end{equation}
which satisfies the original ZRP condition with inverse scattering
$\alpha$. Repeating transformation with $\phi=\exp(-\alpha r)/r$
we obtain $\psi^{(2)}=\psi$. These example shows a relation
between ZRP and DT.

Thus generalized ZRPs are appear as a result of DT. In this
connexion we can raise a question about subsequent dressing of
ZRP. In particular case then we deal with only one prop function
$\phi$ the Crum formulas correspond to the usual DT:
\begin{equation} \label{DT1}
\psi_l^{(1)}=\hbox{const} \cdot (\psi_l'-s\psi_l), \hspace{8mm}
s=(\ln \phi)',
\end{equation}
\begin{equation} \label{DT2}
u_l^{(1)} \equiv \displaystyle
a_0^{(1)}-\frac{l(l+1)^{\mathstrut^{\displaystyle
\mathstrut}}}{2r^2}= u_l +\frac{1}{r^2} - s',
\end{equation}
where we suppose potential $u_l$ describes ZRP. The functions
$\psi_l$, $\phi$ are solutions of the Schr\"{o}dinger equation
$(\ref{E})$. Note the dressed potential $u_l^{(1)}$ is real for
real prop function $\phi$. The next step in the dressing procedure
is a definition of the free parameters of the solutions $\phi$.
Since potential $u_l(r>0)=0$ the solution $\phi$ can be written as
linear combination of spherical functions
\begin{equation}
\phi=Cn_l(\imath \kappa r) +C_1j_l(\imath \kappa r),
\end{equation}
where $C$, $C_1$, $\kappa$ are parameters. The eqs. $(\ref{DT})$
allow to calculate potential in the range $r>0$, but not at $r=0$!
We suppose DT yields a new zero range potential at $r=0$. In order
to solve this problem consider $\phi$ in the vicinity of zero.
There are two different cases. The spherical function properties
show that in the case C=0 the leading term in $\phi$ is $r^{l}$
and in case C=1 one is $r^{-l-1}$. Therefore dressed coefficient
$a_0^{(1)}$ has the following asymptotic at zero
\begin{equation}
\frac{l(l+1)}{2r^2}+u_l^{(1)}\sim\left\{
\begin{array}{cl}
\frac{(l+1)(l+2)}{2r^2} & \hbox{then \ C=0},\\
\frac{l(l-1)}{2r^2}^{\mathstrut^{\mathstrut^{\mathstrut}}}&
\hbox{then \ C=1}.
\end{array}
\right.
\end{equation}
As regards all other possible cases, it is easy to see that ones
lead to the cited above cases. According to eq. $(\ref{DT2})$
dressed potential $u^{(1)}_l$ decreases as $\exp(-2|\kappa| r)$ at
the infinity. Thus, in general DT introduces short-range core of
centrifugal type (which depends on angular momentum $l$) in the
potential. In this situation the boundary conditions on the
dressed wavefunctions $\psi_l^{(1)}$ require modification. We
think that in general case dressed ZRP conventionally represented
as the boundary condition
\begin{equation}
\left. \frac{(\frac{d}{dr})^{2m+1}\,
r^{m+1}\psi_l^{(1)}}{r^{m+1}\psi_l^{(1)}}\right|_{r=0}=
\hbox{const},
\end{equation}
where $m=l+1$ in case $C=0$, and $m=l-1$ then $C=1$. However,
repeating DT for other values $\kappa$ and combining cases $C=0,1$
we can remove short-range core. In the absence of short-range core
the boundary condition looks like eq. $(\ref{Bound})$. The chain
of $N$-DT leads to new S-matrix poles, which does not depend on
value $C_1$
\begin{equation}
\exp(2\imath \delta_l^{(N)})=\frac{(\alpha_l-\imath k^{2l+1})}
{(\alpha_l+\imath k^{2l+1})}\prod_{m=1}^{N}\frac{(\kappa_m-\imath
k)}{(\kappa_m+\imath k)},
\end{equation}
Thus, we can use Darboux transformation in order to add (or
remove) poles of the $S$-matrix. Changing parameter $C_1$ we
obtain potentials with identical spectrum, called phase
equivalent potentials. Such transformation is also known as
isospectral deformation.

{\bf Example 2}: Simplest case of $l=0$ is a instructive example.
Consider original ZRP at $r=0$ with wave function $(\ref{sss})$.
We can choose the solution $\phi$ as
\begin{equation} \label{propf}
\phi=\frac{\cosh(\kappa r)}{r}.
\end{equation}
This choice corresponds the parameters $C=1$ and $C_1=0$. For
short we omit index $l=0$. The DT $(\ref{DT})$ gives rise to the
following property of the dressed wave function
\begin{equation}
\left.\frac{(r\psi^{(1)})'}{r\psi^{(1)}}\right|_{r=0}=\frac{k^2+\kappa^2}{\alpha},
\end{equation}
which slightly differs from usual boundary condition in ZRP theory
$(\hbox{ln} r\psi)'=-\alpha$. The dressed potential has the
short-range tail:
\begin{equation}
u^{(1)}(r>0)=-\frac{\kappa^2}{\cosh^2(\kappa r)}.
\end{equation}
Our observation shows that some partial values $C_1$ can give a
long-range interaction, which looks like $\sim r^{-2}$.

The model we study describes the scattering of an electron on a
compound particle. There were attempts to account this important
circumstance by matrix potentials to be applied not only to
well-known multichannel problem, but to composite particles as
well \cite{SB}. The matrix is a  projection of a complicated base
that includes the orbital momenta, the only possible place for
which is the potential if one restrict himself by the one-particle
case. One could consider our proposal as an attempt to find some
way in the general multi-particle space, that is especially of
importance in the multicenter problem, to be studied in the next
section.

\section{Dressing in a multi-center problem} \label{S2}
The principal observation allows to built a zero-range potential
eigenfunction in the multi-center problem.
In a more general situation one can consider a system with a smooth potential
plus a number of ZRP. If one knows the Green function for the smooth
potential, then one can provide a solution for the problem with
the ZRPs added. This was outlined in $\cite{DO1975}$, where the
case of a single ZRP was considered. Generalization to the case
with an arbitrary number of ZRP is straightforward. On the
contrary, our general idea is to "dress" a multicenter system
without Green function consideration. This procedure gives simple
formulas for partial phase and their corrections at low energies.

Let us consider scattering problem for a non-spherical potential
$U$:
\begin{equation}\label{EQU3D}
\left( -\frac{1}{2}\frac{\partial^2}{\partial r^2} -\frac{1}{r}
\frac{\partial}{\partial r}+\frac{\hat{L}^2}{2r^2} + \hat{U} -
E\right) \psi(\vec{r})=0,
\end{equation}
where $\hat{L}^2$ is square of angular momentum operator, $E$
describes the energy of particle. The asymptotic of wave function
$\psi(\vec{r})$ looks like
\begin{equation} \label{AS3D}
\psi(\vec{r}) \stackrel{\rm r \rightarrow \infty}{\sim}
\exp(\imath \vec{k}\cdot\vec{r}) + f(\theta) \frac{{\rm e}^{\imath
kr}}{r},
\end{equation}
where $f(\theta)$ is scattering amplitude, which depends on
scattering angle $\theta$. The operator $\hat{L}^2$ commutes with
all radial derivatives, in particular with
$\partial=\partial/\partial r$. In three-dimensional space the DT
may be reduced to one-dimensional matrix (or operator) problem
$(\ref{E0})$ with appropriate chosen variable $x$. In our case
$x=r$, functions $a_2$, $a_1$ are the same as in the section
$\ref{S1}$ and
\begin{equation}
a_0=\frac{\hat{L}^2}{2r^2}+\hat{U}.
\end{equation}
The radial DT for any solution of Schr\"{o}dinger equation is
similar as in the section $\ref{S1}$, but $s$ must be assumed as
function of the operator variable $\hat{L}^2$. The transformation
of the potential can be written as
\begin{equation} \label{DTU3D}
\hat{U} \rightarrow \hat{U}^{(1)} = \hat{U} + \frac{1}{r^2}- s'.
\end{equation}
In order to find operator $s$ we can use covariance principal for
equation $(\ref{EQU3D})$. The covariance principal $(\ref{E1})$
formally yields explicit constraint for $s$, which looks like as
\begin{equation}
\begin{array}{c}
a_0'+[a_0,s]+(a_1s)'+[a_1,s]s+\{a_2(s'+s^2)\}'+[a_2,s](s'+s^2)=\\
a_0'+(a_1s)'+\{a_2(s'+s^2)\}'=0^{\mathstrut^{\mathstrut}}.
\end{array}
\end{equation}
Integrating over $r$ we obtain operator equation for s, which in
our case can be written as
\begin{equation}\label{EQS}
s'+\frac{2}{r}s+s^2=\frac{\hat{L}^2}{r^2}+2\hat{U}+\hbox{const},
\end{equation}
where "const" is some function of the operator variable
$\hat{L}^2$, but one does not depend on r \cite{LeZa}. The sense
of "const" can be understood from the asymptotic behavior of s at
infinity. Suppose, passage to the limit $r$ to infinity gives
$s(\infty)=K(\hat{L}^2)$ then const$=K(\hat{L}^2)^2$. In principal
operator $s$ can be found as series $\sum_{n=0}^{\infty}
s_n\hat{L}^{2n}$ where coefficients $s_n$ depends only on $r$. It
is easy to show the equation leads to recursion relations for
coefficients $s_n$. For example, the first equation in the region
where $U=0$ looks like
\begin{equation}
s_0'+\frac{2}{r}s_0 +s_0^2=K_0,
\end{equation}
where $K_0$ is zero coefficient in $K=\sum_{n=0}^{\infty} K_n
\hat{L}^{2n}$. The formula $(\ref{DTU3D})$ gives non-local (over
angles) potential which depends on $\hat{L}^2$. Thus, we have the
algorithm that determine the operator $s$ and dressed potential
via operator $K$. For our purpose (cross section evaluation) we
need only partial phases or scattering amplitude related to
operator $K$. In order to find the partial phases for dressed
potential we need to apply the DT to wave function. However, we
have one trouble: in general DT modifies the plane wave
$\exp(\imath \vec{k}\cdot\vec{r})$. Thus, DT applied to wave
function $\psi(\vec{r})$ with asymptotic $(\ref{AS3D})$ gives an
another asymptotic. In some particular cases, special choose of
the operator $K$ allows to avoid one problem. Indeed, consider the
partial wave asymptotic for a non-spherical potential
$\cite{DemRud}$
\begin{equation} \label{as}
\psi_J(\vec{r}) \sim \frac{1}{{2\imath kr}}(e^{\imath kr+\imath
\delta_J} \Lambda_J(\vec{n})- e^{-\imath kr-\imath \delta_J}
\Lambda_J(-\vec{n})),
\end{equation}
where $\vec{n}$ is unit vector directed as $\vec{r}$, $\delta_J$
denote partial shifts, and $\Lambda_J(\vec{n})$ are normalized
eigenvectors of S-matrix operator (partial harmonics). The most
simple formulas for the shifts $\delta^{(1)}_J$ for the potential
$\hat{U}^{(1)}$ result when partial harmonic $\Lambda_J$ are also
eigenvectors of operator $K$. For example, suppose all partial
harmonic $\Lambda_J$ are eigenvector of $K$ but only $\Lambda_0$
has nonzero eigenvalue $\kappa$
\begin{equation}
K\Lambda_0(\vec{n})=\kappa \Lambda_0(\vec{n}).
\end{equation}
The asymptotic dressing is reduced to action of the operator
$\partial -K$ on asymptotic $(\ref{as})$. It is easy to show by
using expression
\begin{equation}
\ln\left( \frac{\kappa-{\rm i}k}{\kappa+{\rm i}k} \right)=-2{\rm
i}\arctan(k/\kappa),
\end{equation}
for real-valued variables $k$, $\kappa$, that DT changes only one
partial shift $\delta_0$ as
\begin{equation} \label{delta}
\delta_{0}^{(1)}=\delta_{0}-\arctan(k/\kappa).
\end{equation}
In this special case we add only one additional parameter. In the
region $k\gg |\kappa|$ the second term of the equation
$(\ref{delta})$ practically does not contribute to the partial
cross section
\begin{equation} \label{parics}
\sigma_J=\frac{4\pi}{k^2} \sin^2 \delta_J.
\end{equation}
One makes an important contribution to the cross section when
$k\approx |\kappa|$ and so one can be considered as correction at
low energies.

In general case DT modifies all partial harmonics and partial
shifts. DT allows to construct new solvable models with additional
parameters. One of most important problems of solvable models is
the problem of fitting them to some physically meaning parameters.
For example, in discussed case the parameter $\kappa$ can be
linked to the effective radius of the interaction or scattering
length. Well known that scattering length can be defined as
derivative $A=-\delta'(k)$ at $k=0$. Considering the equation
$(\ref{delta})$ at low energies we obtain "re-normalized"
scattering length
\begin{equation} \label{IQ}
A^{(1)}=A+\frac{1}{\kappa}.
\end{equation}

\section{X$_{\hbox{n}}$ and YX$_{\hbox{n}}$ structures} \label{S4}

For purpose of illustration we consider scattering problem for a
dressed multi-center potential. The multi-center scattering within
the framework of the ZRP model was investigated by Demkov and
Rudakov $\cite{DemRud}$ (8 centers, cube), Szmytkowski
$\cite{Szmyt}$ (4 centers, regular tetrahedron), and others.

\subsection{Electron-X$_{\hbox{n}}$ scattering problem}

Suppose structure X$_{\hbox{n}}$ contains $n$ identical
scatterers, which involve only $s$ waves. Let $R$ denote distance
between two any scatterers (in points $\vec{r}_m$). There are
three such structures in three-dimensional space - X$_2$, regular
triangle X$_3$, regular tetrahedron X$_4$. The partial waves
$\psi_J(\vec{r})$ and phase shifts can be classified with respect
to symmetry group representation for the structures
X$_{\hbox{n}}$ (n=2,3,4), degeneracy being defined by the
dimension of the representation $\cite{DemRud}$. We use the
partial waves for ordinary ZRPs in general form
\begin{equation} \label{qwer}
\psi(\vec{r})= \sum_{m=1}^n
c_{m}\frac{\sin(k|\vec{r}-\vec{r}_m|+\delta)}{|\vec{r}-\vec{r}_m|}.
\end{equation}
to derive an algebraical equation for partial phase. The s-wave
boundary condition at the points $\vec{r}_m$ leads to an
algebraical problem for a matrix $n\times n$ with compatibility
condition
\begin{equation}
(p+(n-1)q)(p-q)^{n-1}=0,
\end{equation}
where
\begin{equation}
\begin{array}{l}
p=akR + R \tan \delta, \\
q=a(\sin(kR)+\cos(kR) \tan \delta), \\
\end{array}
\end{equation}
From here, it is easy to show that phases satisfy the following
expressions
\begin{equation} \label{phases_Xn}
\tan \delta_J= \left\{
\begin{array}{cl}
-a\frac{kR+(n-1) \sin (kR)}{R + (n-1)a\cos(kR)},\hspace{8mm} & J=0 \\
-a\frac{kR - \sin(kR)^{^{\mathstrut}}}{R - a\cos(kR)},\hspace{8mm}
& J=1,\ldots,n-1.
\end{array}
\right.
\end{equation}
In special case, assuming $n=4$ we obtain the phases of a regular
tetrahedron $\cite{Szmyt}$. The integral cross sections $\sigma$
can be expressed as
\begin{equation}
\sigma=\sigma_{0}+(n-1)\sigma_{1},
\end{equation}
where partial cross sections $\sigma_J$ are given by the equation
$(\ref{parics})$. It is easy to link a scattering length for a
molecule X$_{\hbox{n}}$ and boundary parameter $a$. At large $R$
the parameter $a$ is reduced to a scattering length for an
isolated atom. Starting from equation
\begin{equation}
\delta(k)= -\arctan \left(a\frac{kR+(n-1)\sin(kR)}{R+(n-1)a\,
\cos(kR)}\right).
\end{equation}
we obtain
\begin{equation} \label{link}
A_{\rm X_{\hbox{n}}}= -\delta'(0)=\frac{naR}{R+(n-1)a}.
\end{equation}
Testing the result, plugging $n=1$ gives  $A_{\rm X_{\hbox{n}}}=a$
for arbitrary $R$. The link ($\ref{link}$) defines the monotonic
function, saturated at $a \rightarrow \infty$.

\subsection{Electron-YX$_{\hbox{n}}$ scattering problem}
The structures YX$_{\hbox{n}}$ can be used, for instance, to study
a slow electron scattering by the polyatomic molecules like
$\hbox{H}_2\hbox{O}$, $\hbox{NH}_3$, $\hbox{CH}_4$, etc. For the
sake of simplicity, we suppose that the scatterers X are situated
in vertices of a regular structure X$_{\hbox{n}}$. Let $D$ denotes
the distance between scatterers Y-X and $R$ denotes distance
between scatterers X-X. In this case, position of the scatterer
$\hbox{Y}$ perfectly fixed only if $n=4$ (geometric center of
tetrahedron). And we have the constraint
$R=2\sqrt{\frac{2}{3}}D$. The partial waves can be written as
$(\ref{qwer})$, where the summation should be performed from
$m=0$ to $n$. The partial phases can be derived analytically. The
result is given by the expression
\begin{equation}
\tan \delta_J=-a_{\rm x}\, \frac{kR-\sin(kR)}{R - a_{\rm x}
\cos(kR)}, \ \ \ \ \ J=2,\ldots,n.
\end{equation}
The $t=\tan \delta_{0,1}$ obeys the quadratic equation
\begin{equation}\label{eta12}
\begin{array}{c}
(t + a_{\rm y}k)\left(\frac{t}{n-1}+a_{\rm
x}\left(\frac{k}{n-1}+\frac {\sin(kR)}{R}+t\frac
{\cos(kR)}{R}\right)\right) = \\
\frac{n}{n-1}a_{\rm x} a_{\rm y}\left(\frac {\sin(kD)}{D}+t\frac
{\cos(kD)}{D}\right)^{2^{\mathstrut^{\mathstrut}}},
\end{array}
\end{equation}
where $a_{\rm x}$, $a_{\rm y}$ denote boundary parameters. For
large distances we can interpret ones parameters as scattering
lengths of isolated atoms. Thus, in the limiting case when the
distance $D$ is very large, the expression for $\tan \delta_{0}$
passes to first equation of $(\ref{phases_Xn})$ and $\tan
\delta_{1}\sim -a_{\rm y}k$. This situation corresponds to
independent scattering on a molecule X$_{\hbox{n}}$ and an atom Y.
The substitution $a_{\rm y}=0$ also reduces the $\tan \delta_0$
for structure YX$_{\hbox{n}}$ to $\tan \delta_0$ for structure
X$_{\hbox{n}}$.

Substitution $t= -Ak$, where $A$ denotes the scattering length for
a molecule YX$_{\hbox{n}}$, and passage to the limit $k= 0$ in
(\ref{eta12}) gives the quadratic equation, with the roots $A=0$
and
\begin{equation}\label{root}
A_{\rm YX_{\hbox{n}}}=D\frac{(a_{\rm y}+na_{\rm x})RD+a_{\rm
x}a_{\rm y}((n-1)D-2nR)}{(R +(n-1)a_{\rm x})D^2-na_{\rm x}a_{\rm
y}R}.
\end{equation}
The last root gives monotonic function of the atomic length
$a_{\rm x}$ with the same features as in the previous section. We
believe that the scattering lengths for isolated atoms do not
change much if the atoms form an polyatomic molecule.

The DT discussed in Section $\ref{S2}$ allows to correct cross
sections at low energies. Thus, using the formulas $(\ref{IQ})$
and $(\ref{root})$ we obtain "re-normalized" scattering length for
a molecule YX$_{\hbox{n}}$.

\begin{figure}
\centering
\includegraphics[scale=1.0]{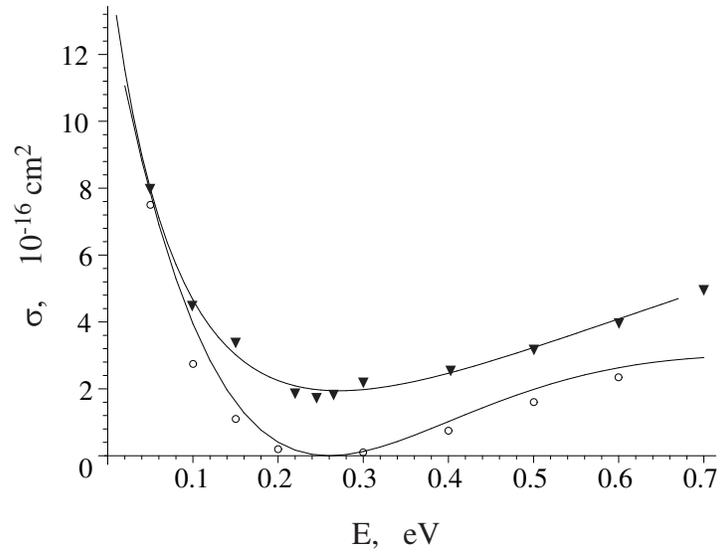}
\parbox[t]{1.0\textwidth}{
\caption{Integral cross sections for electron-Silane scattering
around the Ramsauer-Townsend minimum; upper line - least squares
fitting to the experiment $\cite{Wan}$ (triangles); lower curve
describes our model calculation for partial wave ${\rm A}_1$; open
circles denotes calculation for partial wave ${\rm A}_1$
$\cite{Jain}$} \label{Rmin}}
\end{figure}

\section{Discussion} \label{S5}
In this work we have presented detailed description of the new
solvable models for low-energy electron-polyatomic system
scattering. Now we compare our model calculations with other
theoretical and experimental data. Among all possible applications
we will discuss the scattering by tetrahedral molecule $\rm
SiH_{\rm 4}$ (Silane) because one has most interesting point-group
-- symmetry group of tetrahedron. Let us focus our attention on
one distinct feature of the integral cross section --
Ramsauer-Townsend minimum around $\sim 0.28$ eV.

The authors of the work $\cite{Jain}$ classify the minimum as due
to s-wave scattering into ${^2A_1}$ symmetry and attributes the
main contribution to the cross section at the minimum by the
p-wave scattering via $T_2$ component. Also they write that
minimum is a result of balance of the attractive long-range and
repulsive short-range interactions.

We have performed model calculation and show that
Ramsauer-Townsend minimum is appeared also in consequence of
balance of the attractive short-range and zero-range interactions.
Our calculation is based on the formulas $(\ref{delta})$ and
$(\ref{eta12})$.

In model calculation we used the following parameters (in atomic
units)
$$
\begin{array}{lll}
a_{\rm x}=4.10, & R=4.51, & \kappa=0.185,  \\
a_{\rm y}=1.88, & D=2.762, & \\
\end{array}
$$
which are regarded as constant in the range of interest. The
equilibrium distances $R$, $D$ were taken from ab initio
calculation. The other parameters were chosen so as to reproduce
the realistic low energy asymptotic of $\sigma$ and position of
the minimum. The result of our calculation (lower curve) is shown
in Figure $\ref{Rmin}$. The open circles show numerical
calculation $\cite{Jain}$, triangles and upper line (least squares
fitting) describes the experiment $\cite{Wan}$. Our investigation
shows that "dressing" leads to additional finite range attractive
interaction, which algebraically increases the partial phase
$({\delta_0<0})$ for partial wave ${\rm A}_1$ for ${\rm YX}_4$
structure, and causes to the deep minimum near $0.35$ eV. Thus,
our partial cross sections coincide well with results other
numerical data and coincide in shape with experimental data.

\section{Conclusion}
We introduced a class of models for electron-molecule scattering
description. The principal and novel feature of the model is the
dependence of effective potential on electron momentum (spherical
part of Laplacian). This way we obtain more rich dependence of the
scattering parameters  on k, that improve a coincidence with
experiment in the small energy region. It could be considered as
an alternative to Demkov - Rudakov approach, with generalized
partial waves introduced in each step of dressing procedure.

We write the algebraical expressions for phases of
electron-X$_{\hbox{n}}$ (and -YX$_{\hbox{n}}$) scattering problem.
We hope one can be useful to study a slow electron scattering by a
molecule. We also obtain expressions for scattering lengths, which
probably can be helpful for fitting of parameters (if scattering
length is known). In our calculation we don't use scattering
lengths of isolated atoms, because think that boundary parameters
may differ from ones.

Among the most important aspects of the paper is the demonstration
of  DT power as applied to multi-center scattering problem and ZRP
theory. We established that ZRP can be introduced by DT. Also,
these transformations allow to correct the ZRP model at low
energies. As a novelty we do not use the known solution of the
generalized Miura equation by eigenfuncions of the spectral matrix
problem under consideration (see the Introduction) but construct
particular solutions by means of operator series.

\appendix
\section{Spherical functions properties} \label{APPENDIX}
Below we display notations and some properties of functions used
in the paper. The spherical functions $j_l$, $n_l$ are related to
usual Bessel functions with half-integer indexes $\cite{AS}$.
Ones obey the asymptotic at infinity:
\begin{equation}
j_l(kr)\sim \frac{\sin(kr-\frac{l\pi}{2})}{kr}, \hspace{10mm}
n_l(kr)\sim \frac{\cos(kr-\frac{l\pi}{2})}{kr}.
\end{equation}
For our purpose, it is important that in the vicinity of zero the
spherical functions have the asymptotic behavior at zero:
\begin{equation}
j_l(kr)\sim \frac{(kr)^l}{(2l+1)!!}, \hspace{10mm} n_l(kr)\sim
\frac{(2l-1)!!}{(kr)^{l+1}}.
\end{equation}
Note the double factorial $(2l-1)!!$ satisfies the equation $(2l)!=2^l l!(2l-1)!!$.
Also spherical functions $h_l^{(1,2)}$ appear in our calculations.
Ones link to the functions $j_l$, $n_l$ as
\begin{equation}
\begin{array}{c}
h_l^{(1)}(kr)=n_l(kr)+\imath j_l(kr), \\
h_l^{(2)}(kr)=n_l(kr)-\imath j_l(kr).
\end{array}
\end{equation}
The corresponding asymptotic can be obtained automatically. For
example, at infinity
\begin{equation}
h_l^{(1)}(kr)\sim(-\imath)^l\, \frac{e^{{\rm i}kr}}{kr},
\hspace{10mm} h_l^{(2)}(kr)\sim \imath ^l\, \frac{e^{-{\rm
i}kr}}{kr}.
\end{equation}

\section*{Acknowledgements}
The work is supported by  KBN grant PBZ-Min-008/P03/03. We
acknowledge also the important advices  of
%V. Ostrovsky and
I. Yurova.
% and discussions with J. Sienkiewicz and
%M. Zubek.

\section*{Refferences}

\end{document}